\documentclass[11pt]{article}
\usepackage[margin=1.0in]{geometry}
\usepackage{graphicx}
\usepackage{epsfig}
\usepackage{amssymb,amsfonts}
\usepackage{mathrsfs}
\usepackage{epstopdf}
\usepackage{hyperref,color}
\usepackage{natbib}
\usepackage{setspace}
\usepackage{xypic}\usepackage[all,cmtip]{xy}

\usepackage{braket}
\usepackage{amsmath}

\pdfoutput=1

\begin{document}

\title{Spacetime functionalism from a realist perspective}
\author{Vincent Lam and Christian W\"uthrich\thanks{We wish to thank our audience in Lisbon. We acknowledge support from the Swiss National Science Foundation (grants 105212\_169313 and PP00P1\_170460). VL also acknowledges the Agence Nationale de la Recherche under the  programme `Investissements d'avenir' (ANR-15-IDEX-02) for partial support. We are grateful to Alberto Corti and David Yates for discussions and comments on earlier drafts. We also thank two referees of this journal for their useful comments.}}
\date{23 March 2020}

\maketitle

\begin{center}
\noindent
{\small Forthcoming in \textit{Synthese}.}
\end{center}

\begin{abstract}
\noindent In prior work, we have argued that spacetime functionalism provides tools for clarifying the conceptual difficulties specifically linked to the emergence of spacetime in certain approaches to quantum gravity. We argue in this article that spacetime functionalism in quantum gravity is radically different from other functionalist approaches that have been suggested in quantum mechanics and general relativity: in contrast to these latter cases, it does not compete with purely interpretative alternatives, but is rather intertwined with the physical theorizing itself at the level of quantum gravity. Spacetime functionalism allows one to articulate a coherent realist perspective in the context of quantum gravity, and to relate it to a straightforward realist understanding of general relativity.
\end{abstract}


\section{Introduction: various spacetime functionalisms}
\label{sec:intro}

\emph{Spacetime is as spacetime does} was the slogan at which we arrived in a recent article \citep{LamWuthrich18} in which we argued that spacetime functionalism offers the most promising template to understanding the emergence of spacetime in the context of quantum gravity (QG), the domain of research aiming at a theory of gravity which incorporates both quantum and relativistic effects. The central tenet of spacetime functionalism is that spacetime, qua ontological posit of a theory, is dispensable if the structures postulated by that theory are sufficient to execute all the relevant functions which would otherwise be performed by spacetime itself. Without pretending to be comprehensive, these functions certainly include the determination of spatial distances and temporal durations, and of spatial and temporal relations between physical objects more generally, and so of their relative localisation, i.e., an object's or event's localisation in space and time relative to other, usually nearby, objects and events, which thus furnish a local frame of reference. The project of establishing that the fundamental ontology of a theory fulfills all requisite functions of spacetime becomes an urgent one in QG, where many theories assume or entail that the structures described by these theories are non-spatiotemporal in ways which can be made precise. Thus, quantum theories of gravity often do not include anything resembling spacetime in their ontology. Given that our world is manifestly spatiotemporal, these theories must therefore explain this manifest appearance if the fundamental ontology is non-spatiotemporal. 

Functionalism about spacetime entails that in order to completely discharge this task, quantum theories of gravity need not give rise to spacetime in its full glory. In other words, not all aspects of a relativistic spacetime must be recovered from a more fundamental theory; it suffices that the `functionally relevant' ones are, i.e., those which contribute to the essential functions of spacetime, such as specifying relative localisation, spatial distances, or temporal durations. Moreover, functionalism insists, no purely introspectively accessible, irreducible phenomenal spacetime qualities---`spacetime qualia'---need be accounted for.\footnote{If one wishes to pursue the issue of `spacetime qualia', similarly to the context of philosophy of mind (where it is sometimes called the `hard problem'), then various further standard specifications of functionalism can be invoked in order to solve the issue (see \citeauthor{LeBihan2020} forthcoming). However, we are doubtful that the issue is a genuine one in the spacetime context in the first place, as we have argued in (reference blinded).}

In the predecessor paper, we showed how relative localisation can be functionalised in loop quantum gravity and sketched how spatial distance may be functionally reduced in causal set theory.\footnote{Although we follow common parlance in quantum gravity to speak of spacetime `emergence', it should be noted that although the higher-level entity---spacetime---displays qualitatively novel features, the situation is consistent with a broadly reductive relationship between the fundamental theories of quantum gravity and general relativity. Thus, we accept, as is common in physics and in philosophy of physics, that emergence and reduction are not mutually exclusive. In this sense, the emergence at stake is `weak' rather than `strong' in the distinction popular in philosophy.} Thus, we argued that spacetime functionalism gives us---at least currently---the best framework in which not only to address the challenge of the emergence of spacetime in QG, but also of articulating what kind of response the challenge demands in the first place. 

Spacetime functionalism in QG is particularly attractive for scientific realists, who adopt a ``positive epistemic attitude toward the content of our best scientific theories and models, recommending belief in both observable and unobservable aspects of the world described by the sciences'', as a standard reference defines it \citep{Chakravartty2017}. Such an attitude may be considered premature in QG, where theories remain less than fully developed and lack empirical support of the kind normally required before realism recommends belief in a theory, lock, stock, and barrel. We agree that it would be premature to assume a realist attitude in QG, but we are interested in developing realist strategies \textit{in case such a non-spatiotemporal theory will be borne out}. It is this conditional claim that we are interested in for the purposes of this article. Below, we will dub this stance `presuppositional scientific realism'. Under the presupposition of a fundamental non-spatiotemporal theory, the realist would have to accept its non-spatiotemporal ontology and would inevitably be tasked with demonstrating the emergence of spacetime. It is in this situation that we believe spacetime functionalism would become an essential tool. 

QG is not the only context in which a form of spacetime functionalism has been invoked. In non-relativistic quantum mechanics (QM), it has been used as a template for wave function monism to recover the spatial (though not the temporal) aspects of the world. This was necessitated by wave function monism's tenet that only the wave function in its configuration space fundamentally exists, with physical space emerging from the fundamental structure.\footnote{Presumably, time in some form or other is also in the wave function monist's ontology.} Second, in general relativity (GR), spacetime functionalism has been defended as a successor or a development of the dynamical approach to understanding GR and the role of spacetime in it. On this approach, it is a mistake to hypostatise spacetime already at the level of classical GR, which gets reinterpreted as being committed to much less than the full spacetime structure. Just as in the case of QG and non-relativistic QM, the advocate of spacetime functionalism in GR must show how this reduced ontology suffices to have all pertinent functions of spacetime be executed. 

Spacetime functionalism, at the most abstract level, `functionalizes' spacetime and spatiotemporal properties. This is accomplished first by understanding spacetime and spatiotemporal properties in terms of their functional roles, such as spacetime localization, or spatial distances. Second, it is then shown how the more minimal ontology suffices to fill these functional roles. Hence, as the spacetime functionalisms in the three contexts are all tasked with recovering the relevant functions attributed to spacetime from more minimal structures, all three forms of spacetime functionalism are properly so called.

However, there are also significant ways in which the three cases come apart, or so we wish to argue. As defenders of spacetime functionalism in the context of QG who remain critical of spacetime functionalism in GR (and of the dynamical approach more generally), we have been accused of something close to being inconsistent.\footnote{We are not aware of being so charged in published work, only in discussions.} The charge seems to be this: given that GR is a predecessor to any quantum theory of gravity, and given that a succeeding quantum theory of gravity may well deny spacetime in its ontology, a spacetime anti-realist in QG (who is motivated by her spacetime anti-realism to endorse spacetime functionalism in QG) ought to discount an ontologically robust status of spacetime already at the level of GR. Thus, something like the threat of the pessimistic meta-induction over successive theories which disagree about the ontological status of spacetime should caution us against adopting a prematurely realist attitude towards spacetime in earlier, less fundamental theories. 

The situation vis-\`a-vis wave function monism is somewhat different. First, it differs in that wave function monism responds to the more desperate ontological plight arising from the quantum measurement problem and from the deep puzzle of quantum non-locality. Spacetime functionalism in GR does not respond to a remotely similar interpretative distress. Second, whether or not the pessimistic meta-induction gets any traction in the case of wave function monism depends on how one conceives of the relationship between non-gravitational quantum physics with its implicit assumptions regarding the nature (and existence!) of spacetime and QG.

However, even if a quantum theory of gravity \textit{is} appropriately deemed a successor theory to GR, QM, or both, there is a decisive disanalogy between quantum gravity (QG) on the one hand, and GR and QM on the other. For GR and QM, natural alternatives to those interpretations invoking spacetime functionalism are currently available, whereas, we maintain, this is not the case in QG. For GR, geometric realism about spacetime is the standard option, and varieties of substantivalism or structural realism will fit the realist bill. In QM, primitive ontology approaches presuppose a spacetime arena for their basic existents. Thus, in both cases, the realist has options avoiding spacetime functionalism, options which are simply not on the table when it comes to QG, or so we will argue. As we see it, spacetime functionalism is the realist's only hope in QG.\footnote{At least as things currently stand---see the discussion at the end of section \ref{sec:together}.}

In the end, we want to show how one can be a spacetime functionalist in QG without any attendant commitment to spacetime functionalism either in QM or GR. The central reason allowing us to cleave apart the cases is that the inverse pessimistic meta-induction from spacetime anti-realism in QG to spacetime anti-realism in GR in particular can be blocked. Spacetime anti-realism may be our only option in QG, whereas in QM or GR, it is definitely not. 

We will proceed by accepting two basic and related assumptions for which we will not argue here. The first is what we would think is a reasonably weak form of naturalism, which asserts that philosophical theorizing is continuous with the scientific investigation of the world, and so must at least be consistent with our best science. Consequently, we accept that successful science is our best guide to a comprehensive metaphysics of the world we inhabit. Accepting this naturalism in no way entails that even a well-confirmed scientific theory is not subject to further philosophical and scientific scrutiny and is immune from being rejected; but it \textit{does} mean that in order to contradict an empirically well-confirmed theory, one must provide correspondingly convincing reasons. The better confirmed the theory, the stronger these reasons must be.

Our second basic tenet is a form of scientific realism, or rather a \textit{presuppositional} scientific realism. Scientific realism admonishes belief that our best scientific theories at least approximate a true representation of the natural world as it is. Importantly, this includes unobservable aspects these theories ascribe to the world. Given the speculative and unconfirmed status of theories of QG, adopting a realist attitude would clearly be premature. The presuppositional version of scientific realism deems the interpretation of scientific theories a central task of philosophy (of physics). Accepting presuppositional scientific realism thus enjoins asking what the world would be like if the theory at stake \textit{were} true. The present study sets out from a scientific realist perspective, or at least a presuppositional scientific realist perspective, as may be appropriate. 

In section \ref{sec:qm}, we will start with a discussion of spacetime functionalism in QM, arguing that it directly competes with a `matter-in-spacetime' or `local beables' assumption that is legitimate in this context since external to the scope of the theory. Similarly, in section \ref{sec:gr}, we will see that, within the framework of GR, spacetime functionalism faces the straightforward alternative of a form of (moderate) structural realist understanding of spacetime. Against this background, we will highlight in section \ref{sec:qg} the specific status of spacetime functionalism in QG. We consider spacetime functionalism in a realist perspective in section \ref{sec:together}, with a focus on the inter-theoretic relationship between QG and GR. The main conclusions are drawn in section \ref{sec:conc}.

\section{Functionalism in quantum mechanics}
\label{sec:qm}

The quantum wave function is a central element of the extremely successful theory of non-relativistic quantum mechanics (QM), particularly in its realist formulations such as the Everettian or many-worlds theories (MW), Bohmian mechanics (BM) and dynamical collapse theories, whose most well-known representative is the theory of Ghirardi, Rimini and Weber (GRW). As a consequence, in a standard scientific realist perspective on QM---in either the MW, BM or GRW formulations---it is natural to take the wave function as encoding objective physical features of the world. This realist position is known as the `$\psi$-ontic' view. For any advocate of a $\psi$-ontic view, the exact nature of the wave function is therefore a crucial issue in the debate on the ontology of QM.

Let us make two remarks before we proceed. First, there are at least two ways to interpret the wave function and its import in ways which deny that it represents a real physical feature of the world. These `$\psi$-epistemic' views conceive of the wave function as representing either collective properties of an ensemble of systems or some agent's rational beliefs or information about a quantum system, as does the recently popular `QBism'. What follows will not be relevant for $\psi$-epistemic views.\footnote{See \citet[Ch.\ 3]{Maudlin2019} for an argument against $\psi$-epistemic views based on the so-called `PBR theorem' by \citet{PBR}.} Second, a wave function is of course a mathematical object, typically a complex-valued square integrable function (possibly satisfying certain additional conditions), e.g.\ $\psi \in L^2(\mathbb{R}^{3N})$. For the sake of readability, we will use the same term `wave function' to denote the objective physical features of the world described by this mathematical object. 

A straightforward way to implement the standard scientific realist commitments is to take the quantum wave function to be a concrete physical object, namely a concrete physical field. It is clear that the wave function can also be understood in terms of other metaphysical categories, for instance in terms of properties and relations or in nomological terms or in terms of an altogether new category (as suggested in \citealp[89-90]{Maudlin2019}). But understanding the wave function as a concrete physical field is the most straightforward realist interpretation, indeed suggested by its very mathematical representation, very much like the most straightforward realist perspective on (classical) electrodynamics takes the Maxwell field to be a concrete physical field. But it is well known that there is a crucial difference between a classical field such as the Maxwell field and the quantum wave function: whereas the former is defined over ordinary 3-dimensional space (or 4-dimensional spacetime), the latter is defined on $3N$-dimensional configuration space of QM, where $N$ is the number of particles under consideration, ultimately the total number of particles in the universe. 

At this juncture, there are some interpretative decisions that need to be taken. First, what kind of thing is the entity described by the wave function? A straightforward way to proceed is to reify configuration space and claim it contains the object described by the wave function. David \citet{Wallace2008} has criticised this approach and, together with Chris Timpson, proposed the alternative `spacetime state realism', which asserts that the wave function expresses facts about quantum states associated to spacetime regions \citep{WallaceandTimpson2010}. Let us proceed though assuming the more standard way of a direct reification of configuration space and its denizens. 

The second decision point is whether to admit anything other than the wave function (and the space in which it lives)  into one's ontology. A possible response is `no', and so the resulting ontology for QM does not include 3-dimensional space in the fundamental furniture of the world. It should be noted that such an `ontology without space' results from a rather straightforward---but perhaps too flatfooted---reading of the main realist formulations of QM, namely MW, GRW and even BM. This position is often called `wave function monism'.\footnote{See \citealp[ch.\ 5-6]{Albert2015} and the papers in \citealp{NeyandAlbert2013}. The label `wave function realism' is also frequently used in the literature, but `wave function monism' more specifically underlines the fact that the quantum ontology is only about the wave function on configuration space and nothing else (a single material point in configuration space, the so-called `marvelous point', can be added within the framework of BM, see \citealp[ch.\ 6]{Albert2015}).} 

An obvious issue arises in the context of wave function monism thus defined: how do we recover the familiar picture of 3-dimensional objects located in 3-dimensional physical space? The familiar 3-dimensional picture is crucial to the very empirical evidence for physics in general and for QM in particular: indeed, it seems that any measurement results always ultimately involve 3-dimensional objects such as a measurement apparatus located in 3-dimensional space. David Albert argues that a functionalist perspective can successfully address this issue (\citealp[ch.\ 6-7]{Albert2015}): the claim is that all the objects located in ordinary 3-dimensional space can actually be understood in terms of the causal relations they enter into, that is, in terms of their causal or functional roles, and that the dynamics of the wave function in $3N$-dimensional (configuration) space (as encoded in the Hamiltonian) is such that it can play these 3-dimensional roles---including their spatial and geometrical aspects---that is, instantiating the right sort of causal relations. The details of the proposal are not needed here. But we want to point out two important things for our purpose. 

First, functionalism suggests here a strategy for ordinary 3-dimensional configurations to emerge from an ontology that does not include 3-dimensional space among its basic elements. Despite the differences,\footnote{The differences between the contexts of the ontology of QM and the emergence of spacetime in QG---in particular regarding time---have been discussed in (reference blinded).%to be replaced once accepted
} this provides a natural inspiration for the issue of the emergence of spacetime in QG. 

Second, the ontology of wave function monism, without 3-dimensional space as basic element, faces alternative and competing accounts of quantum ontology, such as the `primitive ontology' account where ordinary 3-dimensional space---together with some local material stuff, the local beables---is postulated from the start, rendering Albert's functionalist strategy unnecessary.\footnote{See \citealp{Maudlin2019}.} Various spacetime-based ontologies with different types of local beables (that is, different primitive ontologies) can be specified within the framework of the main realist formulations of QM, namely, within GRW (flash or matter density ontologies), BM (point particles ontology) and even, to some extent, MW (matter density ontology). Furthermore, and as was already mentioned, Wallace and Timpson's spacetime state realism evidently avoids the need for spacetime functionalism. This is not the place to discuss the benefits and difficulties of wave function monism vis-\`a-vis ontologies of local beables or spacetime states. But it should be clear that, from a scientific realist point of view, there is nothing illegitimate as such in the context of QM about postulating local beables on 3-dimensional space (or in 4-dimensional spacetime). 

Indeed, on the one hand, a scientific realist attitude towards a physical theory requires a specification of what this theory refers to in the world; that is, what one is realist about with respect to this theory. In yet other terms, the scientific realist is required to specify an ontology for the theory. On the other hand, in general, such an ontology cannot be directly and unambiguously read off from the mathematical formulation of the theory,\footnote{But sometimes the situation is rather straightforward, as exemplified in the case of wave function monism.} since formulating a physical theory always involves some metaphysical presuppositions. This latter point has been made clear in the context of the debate on the appropriate methodology for developing a naturalistic or scientific approach to metaphysics and ontology: in particular, it has been convincingly argued that there is no direct entailment from science, and physics more specifically, to metaphysics and ontology (see \citealp{Chakravartty2013}). The two domains are best understood as being intertwined, and as being responsive to each other.\footnote{This attitude is nicely captured by \citet[48]{Callender2011}: ``In slogan form, my claim is that metaphysics is best when informed by good science, and science is best when informed by good metaphysics.'' How exactly science and metaphysics should engage with one another is of course a difficult issue that goes beyond the scope of this paper.\label{ftn:Callender}} In this perspective, what is important to highlight here is that, in the context of QM, postulating a background 3-dimensional space endowed with a certain type of local beables (point particles, flashes, matter density field, \ldots) or states associated with 4-dimensional regions of spacetime is perfectly legitimate, since space (and spacetime) do not belong to the domain described by standard QM (space is not considered as a dynamical entity in any way within standard QM). That this `matter-in-spacetime' assumption is `external' to QM is in line with the fact that the various primitive ontologies for GRW, BM and MW do not lead to any predictive difference from their wave function monistic counterparts. QM itself is silent on the issue.\footnote{This does not exclude the fact that GRW, BM and MW---while all accounting for the empirical success of the QM predictive algorithm---can themselves lead to certain different predictions in principle.}

As a consequence, the functionalist strategy to recover ordinary 3-dimensional situations and their geometrical features from the dynamics of the quantum wave function evolving in $3N$-dimensional space directly competes at the interpretational level with the spacetime-based quantum ontologies of local beables. In this context, the latter can be argued to be less revisionary (in the primitive ontology perspective, all physics, be it classical or quantum, is about matter in spacetime, or local beables evolving in time), and functionalism about (objects in) 3-dimensional space can be seen as a contrived philosophical move to save a counter-intuitive ontology. Consequently, spacetime functionalism is not even mildly mandatory. As we will see in section \ref{sec:qg}, this is in stark contrast to the situation in QG. But before that, we now turn to the other ingredient for QG besides quantum theory, namely GR, which has also been argued to be best understood along functionalist lines.  

\section{Functionalism in general relativity}
\label{sec:gr}

Moving to the object of our analysis---spacetime---the traditional debate between substantivalists and relationalists is a debate over whether to adopt realism about spacetime as an independent entity. Spacetime substantivalism takes spacetime to exist qua independent structure, i.e., to exist as a particular in its own right, over and above the material content of the world. In contrast, relationalism denies that spacetime exists as an independent substance, accepting only the material content of the world into its fundamental ontology. Spacetime, for the relationalist, is merely defined through spatiotemporal relations among the material objects of the world. On the related view that has become known as the `dynamical view', spacetime is, as glorious as it may be, a ``non-entity'', as \citet{bropoo06} argue at least at the level of special relativity (SR).\footnote{The case for the dynamical approach is more convincing for SR than for GR; at best, it seems to collapse into the geometric approach in the context of GR (\citeauthor{read19} forthcoming).} The dynamical approach is motivated by a search for a deeper explanation of the `chronogeometrical significance' of the metric field and is contrasted with the more orthodox `geometric' view. 

As naturalists, let us turn to GR as our best spacetime theory in order to progress on this debate. In standard presentations of GR, a spacetime is represented by an ordered pair $\langle M, g_{ab}\rangle$, where $M$ is a four-dimensional continuum manifold of pointlike events and $g_{ab}$ is a metric field defined on it. Realism about GR would seem to mandate realism about the object described by $\langle M, g_{ab}\rangle$. But of course there is a vigorous debate over the interpretation of GR and over the status of $\langle M, g_{ab}\rangle$ within the naturalist camp. In fact, these debates bring to light that substantivalism and relationalism are both under severe pressure in GR. So let us unpack various approaches to interpret GR a bit.

On the standard geometric view, spacetime, or more specifically the metric field $g_{ab}$, is an autonomous physical entity. Its function is to constrain the dynamics of the matter field present in the universe. Although rarely explicitly articulated, one can take this standard view as being a form of spacetime substantivalism, or at least some sort of realism about spacetime and its geometry. In contrast, the dynamical view as developed by \citet{bropoo06} and \citet{brown2005} takes spacetime and its geometry to merely encode the symmetries of the dynamics of the attendant matter fields. Thus, it is at least closely related to a traditional form of relationalism, according to which spacetime is nothing but the complex of spatiotemporal relations among material bodies. 

At the level of SR, whether one sides with the dynamical or the geometric view hangs to a significant degree on what one takes to stand in need of explanation or, conversely, what one stipulates as primitive. At the level of GR, however, the dynamical view is no longer tenable, at least as originally conceived: because the gravitational field has its own degrees of freedom, it is simply no longer the case that spacetime can merely be the codification of the dynamical properties of matter fields. The geometry of spacetime does not supervene on the dynamics of the matter content of the universe, or any of the matter fields' properties for that matter. It affords physically meaningful distinctions not reflected in the dynamics of matter fields: geometry, as it were, cuts more finely than dynamics. Of course, \citet{brown2005} is aware that ``the situation is more complicated in GR'' (177) than it is in SR. He believes that the source of this difference lies in the fact that GR assumes matter to be classical and closes the book by suggesting that the dynamical view thus reinforces the importance of obtaining a quantum theory of gravity. Needless to say, this is not how we see the situation, although we agree that matters are more involved at the level of GR that they were in SR. 

This priority of geometry over dynamics in GR thus suggests that spacetime and its geometry assume the status of independent entity, and so the dynamical view in its original form becomes untenable. Of course, this does not mean that the original motivation for the dynamical view---to articulate an explanation of the chronogeometric significance of the metric field---evaporates in the context of GR. But if this is an explanatory itch, then it will have to be scratched by invoking resources beyond the matter dynamics and its symmetry properties. 

One way to develop the dynamical view in the context of GR is to transform it into a form of spacetime functionalism. Eleanor \citet{knox2013,knox2014} has done just that. For Knox, something is spacetime ``just in case it describes the structure of the inertial frames, and the coordinate systems associated with these'' (\citealp[15]{knox2014}), ``where inertial frames are those in whose coordinates the laws governing interactions take a simple form (that is universal insofar as curvature may be ignored), and with respect to which free bodies move with constant velocity'' \citep[\S4]{Knox2019}. Given inertial frame functionalism's reliance on elementary concepts of a force-free body and of moving at constant velocities, at least some primitive aspects of spacetime are presupposed. This is noted also by \citeauthor{Esfeldforth} (forthcoming, \S1), who concludes that Knox's functionalism thus only concerns the specific geometry of spacetime and remains within the bounds of standard functionalism, which operates by identifying causal or functional roles of complexes consisting in spatiotemporally related objects. 

Just as in our spacetime functionalism, inertial frame functionalism aims at recovering spacetime structure as it appears in our best theories, and not some kind of `phenomenological spacetime', i.e., the spatiotemporal organization of our experience of the world. Spacetime functionalism, on both Knox's and our version, offers a road to a functional reduction of spacetime as it appears in GR. Her inertial frame functionalism is a `realizer' form of functionalism, since it identifies spacetime with the entity which instantiates the relevant functional role, i.e., the role of defining a structure of inertial frames. This realizer functionalism about spacetime stands in contrast with `role' spacetime functionalism, which identifies spacetime with the property of having certain properties that occupy the spacetime role.\footnote{For more on this distinction as it pertains to spacetime functionalism, see \citeauthor{Yates2018} (forthcoming) and \citeauthor{LeBihan2020} (forthcoming).} Realizer functionalism is a form of realism about the realizer. In the case of Knox, as she freely admits, in the context of GR, this means that the inertial frame functionalist is a realist about the metric field, as it is this field which is responsible for defining a structure of inertial frames in GR. Although motivated by the dynamical view, when it comes to GR inertial frame functionalism thus ends up in a rather different place from the dynamical view, at least as it was originally articulated in the context of SR: it is a solidly realist position about the metric field in GR, and as such just as vulnerable to the pessimistic meta-induction as our preferred kind of realism about spacetime, to be discussed shortly.\footnote{Knox comments on this tension in her recent \cite{Knox2019}.} 

Although relationalists are generally realists about spatiotemporal relations and so about spacetime \textit{structure}, relationalism is a form of antirealism about spacetime (or parts or points of spacetime) \textit{qua independent entity}. In GR, forms of relationalism and so of antirealism about spacetime (or about the metric field) in this sense thus encounter the serious difficulty of accommodating the degrees of freedom of the gravitational field as exemplified, for instance, in the existence of various distinct vacuum spacetimes.\footnote{Of course, as a referee has pressed upon us, the relationalist can circumvent this difficulty by denying that vacuum spacetimes are genuine possible worlds or by claiming that a thoroughgoing empiricist stance makes them uninterested in such worlds. But this response will not carry very far, for two reasons. First, vacuum spacetimes are among the most important spacetimes in GR, scientifically speaking. In fact, many---perhaps most---spacetimes physicists are working on are vacuum spacetimes. If a view on spacetime cannot deal with these exemplar cases, then it can at least not be a candidate for interpreting GR as it is practised by physicists. Second, the fact that there exist gravitational degrees of freedom and that geometry does not supervene on the universe's matter content and its dynamics is completely general and does not depend on the absence of matter. It is just that this fact can be most vividly appreciated in vacuum spacetimes.} This suggests that spacetime \textit{is} an independent entity. However, it is not just antirealism in the sense of relationalism or the dynamic view, which gets under serious pressure. Presuppositional scientific realists must be very careful in choosing their ontological commitments, as illustrated in the infamous hole argument: what is it exactly, in the formalism of GR, that refers, i.e., which describes the physical entity or entities realists should commit to? Without rehashing these arguments here, we take the upshot of this debate to be that a moderate version of ontic structural realism is the most promising form of scientific realism in GR.\footnote{See e.g. \citet{Dorato2000}, \citet{EsfeldandLam2008}, but see also challenges to this view as e.g.\ in \citet{Wuthrich2009}.} Moderate ontic structural realism commits to the existence of relativistic spacetime qua relational structure, i.e., a set of merely structurally identified events standing in spatiotemporal relations to one another. It differs from relationalism in that it can accept that spacetime is an independent entity and from substantivalism in that it denies that spacetime points have an independent, non-structural existence. Given that in GR, the spacetime structure is determined by its topology and its geometry, structural realism is thus closely associated with a more standard geometric view about spacetime according to which, to repeat, spacetime is an autonomous physical entity. 

This geometric structural realism about spacetime thus not only admits relativistic spacetime into its ontology, but also accepts that it is appropriately `spatiotemporal'. In particular, its `spatiotemporality' is an intrinsic feature of this structure and does not merely result from its interaction with matter fields, as the dynamicists would have it. As long as one confines one's attention to GR, we believe that this geometric structuralist realist perspective is superior to any of its competitors. And to so confine one's attention to a scientific theory as successful as GR is natural enough for a scientific realist. Conversely, the dynamical perspective is often assumed as a result of looking beyond GR, typically by requesting an explanation independent of, and prior to, GR of the `chronogeometric significance' of the metric field, which is accepted as a primitive principle in the standard interpretation.\footnote{In this sense, the standard geometric interpretation of GR is an interpretation of \textit{GR}, and there is of course no reason to think that its tenets continue to hold when GR is violated or transcended, i.e., in theories beyond GR.} It is precisely by assuming this perspective from beyond GR that the dynamical view gives up realism, at least about GR, though perhaps not necessarily as a general attitude towards scientific theories. 

At the end of the day it does not matter whether readers agree with our preference for structural realism over its alternatives as an epistemic attitude towards GR. It suffices for present purposes that geometric structuralist realism about spacetime is at least a viable, and arguably an attractive, interpretative option for the presuppositional scientific realist in the context of GR. If that is right, then Knoxian spacetime functionalism about GR has at least a viable competitor.\footnote{A referee suggested that at least intuitively, spacetime functionalism in QM and GR is intra-theoretical, taking a stance on what it is that should be called `spacetime', whereas spacetime functionalism in QG is inter-theoretical, concerned with what recovers spacetime and how. Although intuitive, we do not think that this difference is ultimately tenable: in QM and in GR, spacetime functionalism is as much concerned with a reconstruction project of some antecedently characterized `spacetime' from a more minimal structure. This is why we believe that the cases come apart because this reconstruction project is \emph{optional} in QM and GR, but (presuppositionally, as we admit in section \ref{sec:qg}) mandatory in QG.}

The trouble now arises in the context of QG, where our presuppositional scientific realism (combined with our naturalism) seems to force us into a (presuppositional) antirealism about spacetime. But we are getting ahead of ourselves.

\section{Functionalism in quantum gravity}
\label{sec:qg}

Spacetime functionalism or functionalism about spacetime in QG has a very different status from spacetime functionalism in GR and functionalism in the context of the ontology of QM.\footnote{Even if the functionalist strategy deployed in the context of wave function monism can be a source of inspiration for the QG context to some extent (see section \ref{sec:qm}), the two cases are very different in many ways (see (reference blinded) %to be replaced once accepted
for a discussion of the differences between the two cases).} The issue in the QG context is about the emergence of spacetime from a possibly non-spatiotemporal level. In a previous article \citep{LamWuthrich18}, we have articulated in some detail how a functionalist perspective can alleviate the conceptual difficulties specifically related to the emergence of spacetime from a QG level where it may be absent. In other words, given that spacetime may not be part of the ontological picture sketched by some of the main research programs in QG, functionalism provides conceptual tools to connect such an ontology without spacetime to the ordinary spacetime picture of the rest of classical (and, to some extent, quantum) physics. 

The specific status of functionalism in the QG context (compared to the QM and GR cases) is linked to the status of the claim about the disappearance of spacetime. As it turns out, there are many strong indications that standard spacetime features may just not be there at the QG level, to different degrees in different QG approaches.\footnote{See \citet{HuggettWuthrich2013}, \citet{HuggettVistariniWuthrich}, or \citeauthor{HuggettWuthrich2018} (forthcoming) in the philosophical literature, and \citeauthor{Oritiforth} (forthcoming), \citet{Rovelli2011a}, \citet[especially \S10.1.3]{Rovelli2004}, and \citet{Witten1996} as examples of statements by physicists. We should also note that there has been recent dissent in the philosophical literature by \citet{LinnemannLeBihan}.\label{fn:disappear}} The aim here is not to argue for the disappearance of spacetime in QG; rather, we would like to underline the fact that the issue of the nature of spacetime precisely belongs to the domain investigated by QG, at least as conceived by those approaches to QG taking the background independence and the dynamical nature of the spacetime structure in GR as a starting point for QG theorizing.\footnote{There is a strong case to be made for the emergence of spacetime in the context of string theory as well, as is argued by \citet{Huggett2017}.} In this sense, the claim about the disappearance of spacetime is not `external' to QG, in contrast to the `matter-in-spacetime' assumption in the QM context (see section \ref{sec:qm}). We proceed on the presupposition that spacetime disappears from the fundamental ontology, as is strongly suggested by current research in QG. In this case, we argued in \citet{LamWuthrich18}, spacetime functionalism becomes an indispensable element of understanding how spacetime emerges. 
 
As a consequence, from the point of view of the QG approaches pointing to the disappearance of spacetime---again, to different degrees depending on the considered approaches---assuming a priori an ontological framework for QG relying on some standard smooth spacetime background (e.g.\ assuming a priori an ontology of local beables for QG) is neither physically nor metaphysically legitimate (contrary to what is sometimes claimed in the literature, see \citeauthor{Esfeldforth} forthcoming).\footnote{We insist that the illegitimacy here flows directly from our proclaimed naturalism and so is to be understood \emph{from the point of view of the considered QG approaches that point to the disappearance of spacetime}, for various reasons partly related to the focus on certain physical principles (see \citeauthor{Crowtherforth} forthcoming for the important roles of principles in QG). It does not imply in any way that alternative QG approaches that retain spacetime in one way or another are illegitimate at this stage.} Indeed, in this perspective, such a metaphysical assumption is illegitimate since it directly conflicts with certain physical ingredient principles on which the considered QG approaches are based (such as background independence) and is more generally in tension with the expectation that these QG approaches will shed some new light (hopefully based on experimental evidence!) on the very nature of spacetime. Postulating ontologies on some fixed background spacetime for these QG approaches is therefore clearly not `responsive' to the underlying physics (in terms of Callender's slogan, these ontologies are not `informed' by the relevant physics, see footnote \ref{ftn:Callender}): this stands in direct tension with the naturalism we have adopted and in particular with a naturalistic approach to metaphysics. Moreover, we would like to stress again that an analysis of the most developed approaches to QG clearly shows that all major approaches postulate fundamental structures which are, to some degree and in some way, non-spatiotemporal (see footnote \ref{fn:disappear}).

In this context, spacetime functionalism in QG (or, more precisely, a functionalist perspective on the emergence of spacetime in QG) cannot be merely seen as a contrived philosophical move to save some counter-intuitive ontology, as in the case of wave function monism (see section \ref{sec:qm}). Rather, spacetime functionalism in QG is best considered as a straightforward way to articulate a (presuppositional) scientific realist attitude in the QG context where spacetime or certain standard spacetime features are simply absent. Within this framework of QG without spacetime, the functionalist strategy does not compete with other realist but less revisionary spacetime-based ontologies, since, obviously, these latter are just not available in this case. The specification of spacetime-based ontologies (e.g.\ in terms of local beables) in QG generally amounts to adopting different approaches to QG---approaches that rely on different principles (typically, background independence is eschewed as a foundational principle of QG within the framework of the spacetime-based ontologies for QG, see recently e.g.\ \citealp{Tilloy2018}, \citealp{PintoNetoStruyve2018}). The point here is not to argue in favour of one research program in QG over others, but rather to stress the fact that spacetime functionalism and spacetime-based ontologies cannot be easily compared at the interpretational level in QG since they consider different lines of research that may ultimately lead to (radically) different physical theories based on different physical principles (and which will hopefully be discriminated by experimental evidence).\footnote{This crucial point is overlooked in \citeauthor{Esfeldforth} (forthcoming).} This is in stark contrast to the case of QM discussed in section \ref{sec:qm}, where wave function monism (together with its functionalist strategy to recover standard 3-dimensional configurations) and their counterpart spacetime-based ontologies acknowledge exactly the same empirical evidence (which however has a different ontological status in the two cases, since the underlying ontologies are different). 

So, spacetime functionalism helps to articulate a general ontological framework for QG approaches that do not rely on spacetime being fundamental. However, one might wonder what is the point of considering the non-spatiotemporal ontologies of QG, since these QG approaches are still very far from being the kind of mature physical theories with strong empirical support that would recommend a realist attitude. There are two related points to notice in order to address this worry. First, specifying the ontological picture of some physical theory is not the same as promoting a realist attitude towards this theory---it is the difference between our proposed presuppositional realist stance and standard scientific realism. If the latter requires the former (to some extent), the converse does not hold.\footnote{Indeed, discussing ontological implications of scientific theories may actually lead to a critical attitude towards scientific realism itself (as a good example, see the role of metaphysical underdetermination in QM in \citealp{VanFraassen1991}).} Second, articulating the ontological and metaphysical implications of (fundamental) physical theories can be crucial to the understanding of the theories themselves and to the very process of physical theorizing (for instance, Bell's reflections on locality, determinism and quantum ontology were crucial to his groundbreaking results; those of Einstein on the nature of space and time constitute another paradigmatic example). Similarly, spacetime functionalism ought to be seen as being crucial for the physical theorizing without spacetime, by providing tools for clarifying the conceptual difficulties specifically linked to the emergence of spacetime in certain approaches to QG. 

In forthcoming work, David \citeauthor{Yates2018} challenges the idea that at least a very natural form of spacetime functionalism really helps in the case of the disappearance of spacetime in fundamental QG as we have argued in earlier work \citep{LamWuthrich18}. He poses a dilemma for the realizer form of spacetime functionalism, at least insofar as it denies the existence of non-fundamental spacetime. Realizer functionalism about a theoretical concept normally adopts something like a Kim-type reduction of that theoretical concept to the underlying realizer by means of a Ramsification of the terms naming the concept and its cognates. This remains largely under the surface in Knox's work, but is essentially endorsed by \citet{HuggettWuthrich2013}.\footnote{We remain explicitly uncommitted between realizer and role versions of spacetime functionalism.}

In Yates's dilemma, either the Ramsification is isomorphic to the spacetime propositions it replaces, or it is not. If it is, then the functional reduction succeeds in establishing how propositions involving spacetime concepts can be derived from a non-spatiotemporal fundamental theory, but in this case the fundamental entities can be \textit{identified} with higher-level spatiotemporal entities which were supposed to be absent. Consequently, according to Yates, the fundamental ontology is, after all, spatiotemporal.\footnote{In this case, Yates claims that the problem of empirical coherence is not solved. We would insist, however, that it is \textit{resolved}, as it does not really arise in the first place.} Thus, spacetime functionalism is superfluous on this horn. If the Ramsification is not isomorphic, then the functional reduction sought fails to connect the non-spatiotemporal fundamental theory to higher-level spatiotemporal theories. On this horn then, spacetime functionalism is impotent. On either horn, realizer spacetime functionalism does not do the work it is built to perform. 

So how can this dilemma be resolved? Although, to repeat, we do not wish to commit to a realizer version of spacetime functionalism, we agree that spacetime functionalism had better succeed in connecting the levels---that is the whole point of introducing it. But we disagree that success would mean that the fundamental ontology was spatiotemporal after all, for two reasons. First, judged by their own lights, and still under the presupposition that current research programmes in QG are indicative of future ones, the fundamental theories are radically different from anything spatiotemporal (see again footnote \ref{fn:disappear} for references). Second, models of these theories will in general \textit{not} give rise to spacetime: many of the physically possible ways in which the fundamental existents combine will not yield structures even approximately isomorphic to spacetime (e.g.\ `non-geometric phases' in certain QG approaches). More generally speaking, the two horns of the dilemma are not so cleanly separated in real approaches to QG. The situation is much murkier, and we should expect there sometimes to be substructures at the fundamental level which are (partially) isomorphic to salient aspects of spacetime, and always to be substructures at the fundamental level which are not isomorphic to anything resembling spacetime. This makes the sought functional reduction hard---and the fundamental ontology non-spatiotemporal. As difficult as it may be though, the functional reduction had better succeed for these non-spatiotemporal approaches, given the fact that our world so vividly appears to be spatiotemporal.

\section{Bringing it all together}
\label{sec:together}

Let us return to the alleged tension between our realism about spacetime in GR and the presumed antirealism about spacetime in QG, at least if an approach with a non-spatiotemporal ontology is ultimately borne out. We find ourselves at a dialectical juncture: either we embrace spacetime functionalism as a universal tool applicable in all circumstances, and for all theoretical levels, be it in QM, GR, or QG, or else we resist this universalism and adopt a more piecemeal and local approach, tailored to each situation. We opt for the latter. Let us discuss the options in turn.

Wholesale or universal spacetime functionalism is attractive because it offers a general and widely applicable interpretative template for any theory, particularly for those where the ontological status of spacetime is otherwise precarious. This unified position can be further articulated, for example, in terms of Knox's inertial frame functionalism. If so articulated, and under the additional assumption that we are realists about the realizers of the inertial frame role, it amounts to metric realism in GR, to realism about the relevant aspects of the wave function in QM, and to realism about whatever structures implement, or give rise to, inertial frames in QG. 

If our original motivation was to eschew the pessimistic meta-induction, then we have failed on universal (realizer) spacetime functionalism, as we find ourselves successively committed to structures which later turn out to be obsolete in performing the central role of spacetime, e.g.\ that of inertial frames. Furthermore, universal spacetime functionalism also becomes universally vulnerable to charges that it misidentifies the relevant spacetime function or role. Finally, it implies an inflexibility and so an inability to interpret theories `locally', i.e., on their own terms. One might be worried that this inflexibility stands in tension with a naturalism embracing the freedom of our scientific enquiry of the natural world to reformulate basic principles, to readjust methodological means, and to reconsider the interpretative outlook on its theories. 

Our set of commitments stands in opposition to this wholesale spacetime functionalism. Our retail spacetime functionalism is designed to address a particular set of problems arising from the disappearance of spacetime in the specific context of QG. Of course, it may turn out to be equally applicable in a sufficiently similar context, but there is no presumption that there will be any other sufficiently similar situation. As we have stated above, this piecemeal approach leads ourselves to incline towards a geometric structural realism about spacetime in GR, and spacetime functionalism in much of QG. Obviously, there is no requirement that these `locally optimal' interpretations are consistent across contexts. 

Taking naturalism seriously mandates local interpretations of theories, i.e., their reading needs to start from the theories themselves, rather than from a presupposed and fixed interpretative scheme or set of demands. Given that scientific revolutions may bring with themselves a shift in methods, aims, and values that constitute a scientific paradigm, naturalism prohibits an inflexible a priori commitment to a particular interpretative template. Thus, we believe that scientific realism tempered by naturalism must accept the possibility that our interpretative stances in GR and in QG diverge.\footnote{Of course, as we have discussed in section \ref{sec:qg}, this is also true for the transition from QM to QG, in particular regarding the metaphysical assumption of local beables.} Our stance in favour of local interpretations relies on our naturalism alone, and so is orthogonal to questions regarding spacetime realism/anti-realism or functionalism. 

The local approach to understanding physical theories thus makes plausible that the ontologies of distinct theories differ. Specifically, taking the local approach suggests, at least to us, a geometric structural realism about spacetime in GR. Similarly, it is at least a live possibility that the ontologies of the fundamental theories of QG are non-spatiotemporal. In this sense, the locally optimal, presuppositionally realist interpretation of much of QG recommends a presuppositional antirealism about spacetime. In this latter case, but not in the former, the scientific realist is well advised to opt for spacetime functionalism. Notice that this recommendation is based on the presupposition that a non-spatiotemporal theory of QG is borne out, in which case it would simply be rational to give up realism about spacetime (though of course not scientific realism simpliciter). The functionalism adopted at the more fundamental level recovers how relativistic spacetime is an approximately instantiated concept and so explains why GR was as successful as it was in those domains where it has indeed been successful. In this manner, functionalism offers a way to appreciate the relationship of theories of QG to the earlier, less fundamental theory of GR. 

But doesn't this local approach straddle scientific realism with a set of inconsistent beliefs? It does not, since the commitments arise only in specific contexts and so the resulting beliefs are context-relative: in the context of GR, we may be spacetime realists, and in the context of QG, spacetime antirealists. So there is no inconsistency, but there would of course be a sense in which the spacetime realist in GR would have been wrong: their positive epistemic attitude towards spacetime turned out to be mistaken after all. It is clearly not sufficient for a scientific realist to limit themselves to just interpreting theories locally and refrain from any further judgment; instead, a comparative assessment of different competing theories is also needed. In particular, the scientific realist would have to come to an assessment of different physical theories and so of their epistemic attitudes towards them individually, but also---crucially---of the relationship between these theories. But it is precisely in this task of a detailed evaluation of the relationship between GR and QG that we believe spacetime functionalism will deliver essential services; in fact, it is the only road that we can see to explaining why GR was as successful as it was, given---presuppositionally---that the world is fundamentally non-spatiotemporal.\footnote{This was also the central difference to the case of wave function monism in QM, where viable interpretative alternatives are readily available---something which is currently not the case in QG approaches without spacetime.} Naturally, it would then turn out that the realist about spacetime was mistaken and has to acknowledge the only approximate validity of their spatiotemporal beliefs. But this is just the simple fallibility of science, which is always a risk for the realist; a risk, however, that is exactly the same here as it is elsewhere.

When it comes to interpreting non-spatiotemporal theories of QG, spacetime functionalism is currently the only game in town. However, for all we said here, this conclusion could fail to obtain for two reasons. First, the fundamental ontology could turn out to be fundamentally spatiotemporal, betraying current appearances. In this case, the scientific realist can happily revert to spacetime realism and the functionalism we advocate would become otiose. Second, spacetime anti-realism prevails, but the future reveals alternative, non-functionalist strategies to handle the disappearance of spacetime. Such a discovery would force a philosophical debate concerning the relative merits of these alternatives and spacetime functionalism. We would welcome such a debate, but cannot possibly predict its outcome, given that there is currently but one competitor on the field.\footnote{We thank a referee for giving us the opportunity to clarify the dialectical situation as we see it.}

\section{Conclusions}
\label{sec:conc}

In a previous publication, we have argued that spacetime functionalism provides tools for resolving the conceptual difficulties linked with the emergence of spacetime in QG \citep{LamWuthrich18}. By helping to articulate a coherent picture from some QG level lacking fundamental spatiotemporal features, spacetime functionalism is intertwined with the physical theorizing itself in several approaches to QG eschewing spacetime. We have argued here that it is precisely in this sense that spacetime functionalism in QG is radically different from other functionalist approaches in QM and GR: in these latter cases, the functionalist strategy faces interpretative alternatives (all taken seriously in the philosophy of physics community) in a way that is simply not available in the case of QG approaches without spacetime, at least not currently. In the context of GR, some form of a (moderate) structural realist understanding of relativistic spacetime constitutes an obvious alternative to the inertial frame functionalism of Knox; there is no such straightforward alternative to spacetime functionalism in QG (actually, spacetime functionalism can be understood as part of the most straightforward interpretative strategy in QG, taking at face value the many theoretical indications pointing to the absence of spacetime). In the case of QM, the `matter-in-spacetime' or `local beables' assumption, which is external to the scope of the theory, is an example of a serious contender to Albert's functionalist strategy to recover 3-dimensional situations from a wave function ontology; no such external assumption is available for QG approaches investigating the very fundamental nature of spacetime.   

The crucial point to be stressed is the following: spacetime functionalism is not part of a `spacetime-less' interpretation of QG among other interpretations (as functionalism in QM or GR is); rather it is a way to articulate the physics suggested by several approaches to QG where spacetime is absent. Other approaches to QG may retain spacetime in some more robust way, but their physics will be different, so that the issue about the fundamental existence of spacetime is to be ultimately settled by experiments (hopefully).\footnote{However, one could imagine a case where the issue remains underdetermined, that is, a hypothetical QG theory that can be interpreted either in a spatiotemporal or in a `spacetime-less' way.} As a consequence, spacetime functionalism in QG, in contrast to the functionalist strategies in QM and GR, is not a purely interpretative move and should not be considered as such. If (and only if) naturalism ultimately mandates anti-realism about spacetime in QG, then we see spacetime functionalism as our best hope of recovering the spatiotemporality of our world;\footnote{Of course, spacetime functionalism is not strictly speaking entailed by QG approaches without spacetime and the need to account for the manifest appearance of spatio-temporal features of our world. But spacetime functionalism \emph{is} a convincing way of reconciling the two and of  addressing serious conceptual worries that can be raised this context, as we have extensively argued in our earlier paper \citep{LamWuthrich18}.} in contrast, it is evident that naturalism does not similarly mandate anti-realism in the cases of QM or GR. 

Furthermore, spacetime functionalism provides a clear understanding of the relationship between the QG level without spacetime and the GR level of relativistic spacetime. We have argued that the possible discontinuity of the ontological commitments about spacetime within the frameworks of GR and QG respectively constitutes no threat to the realist, since, in a naturalistic perspective, theories should be interpreted on their own terms, locally so to speak, rather than forced into some presupposed, universal interpretative scheme. Spacetime functionalism precisely allows one to articulate such a realist perspective in the context of QG, and to relate it to a straightforwardly realist understanding of GR.

\bibliographystyle{apa}
\bibliography{Ref}

\end{document}